\documentclass[aps,pra,9pt,twocolumn,showpacs,superscriptaddress]{revtex4-2}

\usepackage{physics}							
\usepackage{latexsym}
\usepackage{amssymb}
\usepackage{graphics,epstopdf}
\usepackage{newlfont}
\usepackage{amsfonts}
\usepackage{epsfig}
\usepackage[colorlinks=true, citecolor=blue, urlcolor=blue]{hyperref}
\usepackage{amsmath}
\usepackage{graphicx}
\usepackage{dcolumn}
\usepackage{bm}
\usepackage{color}
\usepackage{amsthm}
\usepackage{caption}
\usepackage{subcaption}

\newtheorem{theorem}{Theorem}
\newtheorem*{theorem*}{Theorem}

\begin{document}

\title{Stronger Quantum Speed Limit For Mixed Quantum States}

\author{Shrobona Bagchi}
\email{shrobona@kist.re.kr}
\affiliation{Center for Quantum Information, Korea Institute of Science and Technology, Seoul, 02792, Korea}
\author{Dimpi Thakuria}
\email{dimpithakuria@hri.res.in}
\affiliation{Atominstitut, Technische Universität Wien, Stadionallee 2, 1020 Vienna, Austria}
\affiliation{Quantum Information and Computation Group, Harish-Chandra Research Institute, Chhatnag Road, Jhunsi, Allahabad 211019, India and Homi Bhabha National Institute, Anushaktinagar, Training School Complex, Mumbai 400085, India}
\author{Arun Kumar Pati}
\email{akpati@hri.res.in}
\affiliation{Quantum Information and Computation Group, Harish-Chandra Research Institute, Chhatnag Road, Jhunsi, Allahabad 211019, India and Homi Bhabha National Institute, Anushaktinagar, Training School Complex, Mumbai 400085, India}

\begin{abstract} 
We derive a quantum speed limit for mixed quantum states using the stronger uncertainty relation for mixed quantum states and unitary evolution. We also show that this bound can be optimized over different choices of operators for obtaining a better bound. We illustrate this bound with some examples and show its better performance with respect to some earlier bounds.
\end{abstract}

\maketitle

\section {Introduction }\label{intro}
The uncertainty relations are of fundamental importance in quantum mechanics since the birth of quantum mechanics in the early nineties. The uncertainty principle was first proposed by Werner Heisenberg heuristically \cite{Heisenberg1927}. He provided a lower bound to the product of standard deviations of the position and the momentum \cite{Heisenberg1927} of a quantum particle. Not only this, the uncertainty relations are also capable of capturing the intrinsic restrictions in preparation of quantum systems, which are termed as the preparation uncertainty relations \cite{Robertson1929}. In this direction, Robertson formulated the so called preparation  uncertainty relation for two arbitrary quantum-mechanical observables which are generally non-commuting \cite{Robertson1929}. However, the Robertson uncertainty relation do not completely express the incompatibility nature of two non-commuting observables in terms of uncertainty quantification and is not the most optimal nor the most tight one. It also suffers from the triviality problem of uncertainty relations. To improve on these deficiencies, the stronger variations of the uncertainty relations have been proved which capture the notion of incompatibility more efficiently and also provide an improved lower bound on the sum and product of variances of the generally incompatible observables \cite{Pati2014,Mondal2017}. On another note, and along the same lines of formulatio of uncertainty relations, the energy-time uncertainty relation \cite{Aharonov1961,Aharonov2002} proved to be quite different from the preparation uncertainty relations of other observables such as the position and momentum or that of the angular momentum because time is not treated as an operator in quantum mechanics \cite{Busch2008}. Thus, time not being a quantum observable, time-energy uncertainty relation lacked a good interpretation like for those of the other quantum mechanical observables such as position and momentum. Mandelstam and Tamm derived an uncertainty relation \cite{Mandelstam1945} which is now called an energy-time uncertainty relation. It follows from the Robertson uncertainty relation when we consider the initial quantum state and the Hamiltonian as the corresponding quantum mechanical operators \cite{Mandelstam1945} and $\Delta t$ as the time interval between the initial and final state after the evolution. An interpretation of this time energy uncertainty relation was given in terms of the so called quantum speed limit \cite{Aharonov1961,Aharonov2002}. In the current literature, there are several other approaches to obtain quantum speed limits for closed quantum system dynamics \cite{Anandan1990, Levitin2009, Gislason1956, Eberly1973, Bauer1978, Bhattacharyya1983, Leubner1985,  Vaidman1992, Uhlmann1992, Uffink1993, Pfeifer1995,  Horesh1998, AKPati1999, Soderholm1999,  Andrecut2004, Gray2005, Luo2005,  Zielinski2006,  Andrews2007, Yurtsever2010, Fu2010, Zwierz2012, Poggi2013,Kupferman2008, Jones2010, Chau2010, S.Deffner2013, Fung2014, Andersson2014, D.Mondal2016, Mondal2016, S.Deffner2017, Campaioli2018,Giovannetti2004, Batle2005, Borras2006, Zander2007,Ness2022, Pandey22, thakuria2022} as well as for open quantum system dynamics \cite{Deffner2013, Campo2013, Taddei2013, Fung2013, Pires2016, S.Deffner2020,Jing2016,Pintos2021,Mohan,Mohan21,Pandey2022}. Quantum speed limits have also been generalised to the cases of arbitrary evolution of quantum systems \cite{abhay2022}, unitary operator flows \cite{Carabba2022}, change of bases \cite{naseri2022}, and for the cases of arbitrary phase spaces \cite{meng2022}. Most recently, in another direction exact quantum speed limits have also been proposed \cite{Pati2023}.

The notion of quantum speed limit is not only of fundamental importance, but also has many practical applications in quantum information, computation and communication technology. The quantum speed limit bounds have proven to be very useful in quantifying the maximal rate of quantum entropy production \cite{deffner2010,das2018}, the maximal rate of quantum information processing \cite{bekenstein1981,Mohan}, quantum computation \cite{lloyd2000, lloyd2002, AGN12} in optimal control theory \cite{Caneva2009, Campbell2017}, quantum thermometry \cite{Campbell2018} and quantum thermodynamics \cite{Mukhopadhyay2018}. These explorations motivate us to find better quantum speed limit bounds that can go beyond the existing bounds in the literature. In this paper, we use the stronger uncertainty relation developed in \cite{Pati2014}, then generalised to the case of mixed quantum states to derive a stronger form of quantum speed limit for mixed quantum states undergoing unitary evolution. We show that the new bound provides a stronger expression of quantum speed limit compared to the MT like bound for mixed quantum states. This bound can also be optimized over many operators. We then find various examples for mixed states and some example Hamiltonians that shows the better performance of our bound over the MT like bound for mixed quantum states and the bounds for mixed states in Ref.~\cite{Campaioli2018}.

The present article is organised as follows. In sections \ref{QSL} and \ref{SURGMS}, we give the background that includes the various forms of quantum speed limit for mixed quantum states \ref{QSL}, followed by the stronger uncertainty relations for mixed quantum states in \ref{SURGMS}. In section \ref{SQSLUE}, we derive the stronger quantum speed limit for mixed quantum states respectively and show methods to calculate the set of operators obeying a necessary condition for the bound to hold true. In section \ref{examples}, we show its better performance with examples of random Hamiltonians, specific examples of Hamiltonians that are useful in quantum computation, random quantum states respectively over three different previous bounds of quantum speed limit for mixed quantum states . Finally, in Section \ref{discussion} we conclude and point out to future directions.

\section {Background }\label{background}

\subsection {Quantum Speed Limits}\label{QSL}

Quantum speed limit is one of the interpretations of the time energy uncertainty relation in quantum mechanics. In particular Mandelstam and Tamm derived the first expression of the quantum speed limit time as $\tau_{QSL} = \frac{\pi}{2\Delta H}$, where $\Delta H$ is the variance of the Hamiltonian driving the quantum system $H$ \cite{Mandelstam1945}. As an interpretation of their bound, they also argued that $\tau_{QSL}$ quantifies the life time of quantum states. Their interpretation was further solidified by Margolus and Levitin \cite{Margolus1998}, who derived an alternative expression for $\tau_{QSL}$ in terms of the expectation value of the Hamiltonian as $\tau_{QSL} = \frac{\pi}{2\langle H\rangle}$. Eventually, it was also shown that the combined bound,
\begin{align}
\tau_{QSL}=\max\{\frac{\pi\hbar}{2\Delta H},\frac{\pi\hbar}{2\langle H\rangle}\}
\end{align}
is tight. Many more versions of quantum speed limits have been proposed since then, with an intent to improve the previous bounds in terms of tightness and performance. In this direction, recently a stronger quantum speed limit for the pure quantum states has been proposed as follows.
\begin{align}
\tau \geq \frac{\hbar s_0}{2 \Delta H}+\int_{0}^{\tau}R(t)dt,
\end{align}
where we have
\begin{align}
R(t)=\frac{1}{2}|\langle\Psi^\perp(t)|\frac{A}{\Delta A}\pm i\frac{H}{\Delta H} |\Psi(t)\rangle|^2.
\end{align}
The stronger quantum speed limit bound generally performs better than the MT bound for pure quantum states since it can be shown that for pure quantum states $R(t)\geq 0$ in general. On the other hand, quantum speed limits for the mixed quantum states have also been proposed in various forms \cite{Campaioli2018}. Quantum speed limit can be extended to the case of mixed quantum states by defining the distance between the initial state $\rho_0$ and the final state $\rho_t$ as their Bures angle $ {\mathcal {L}} (\rho _0,\rho _t)=\arccos ({\mathcal {F}} (\rho _0,\rho _t))$, with $ {\mathcal {F}} (\rho _0,\rho _t)=\mathrm{tr}[\sqrt{\sqrt{\rho _0}\rho _t\sqrt{\rho _0}}]$ being the Uhlmann root fidelity,
\begin{align}
\label{eqn:mtl}
\tau_{\mathcal{L}}=\frac{\mathcal{L}(\rho_0,\rho_t)}{min\{H,\Delta H\}},
\end{align}
where, $\hbar=1$ has been set for convenience. It bounds the evolution time required to evolve the mixed state $\rho_0$ to the final state $\rho_t$ by means of a unitary operator $U_t$, i.e., $\rho _t=U_t\rho_0 U_t^{\dagger }$, where the quantum system is governed by a time-dependent Hamiltonian $H_t$. There are many other forms of speed limits for mixed quantum states, which we leave for later investigation in future research.
In \cite{Campaioli2018} another bound tighter than the MT bound was derived for the speed of unitary evolution. According to this bound, the minimum time required to evolve from state $\rho$ to state $\sigma$ by means of a unitary operation generated by the Hamiltonian $H_t$ is bounded from below by
\begin{align}
\label{eqn:pre1}
T_{\Theta}(\rho,\sigma)=\tau_2=\frac{\Theta(\rho,\sigma)}{Q_\Theta} ~~~\mathrm{where} \\ 
Q_\Theta=\frac{1}{T}\int_0^T \mathrm{d}t\sqrt{\frac{2\mathrm{Tr}(\rho_t^2H_t^2-(\rho_tH_t)^2)}{\mathrm{Tr}(\rho_t^2-\frac{1}{N^2})}} ~~~\mathrm{and} \\
\Theta(\rho,\sigma)=\arccos\sqrt{\frac{(\mathrm{Tr}(\rho\sigma)-\frac{1}{N})}{(\mathrm{Tr}(\rho^2)-\frac{1}{N})}}\label{better bound}
\end{align}
where $N$ is the dimension of the quantum system undergoing unitary evolution due to the time independent Hamiltonian $H$. We mention this bound since this bound does not reduce to the MT bound in general. However, there is another bound proposed in the same paper that reduces to the MT bound for the case of pure states. It is given as follows 
\begin{align}
\label{eqn:pre2}
T_{\Phi}(\rho,\sigma)=\tau_2=\frac{\Phi(\rho,\sigma)}{Q_\Phi} ~~~\mathrm{where} \\ 
Q_\Phi=\frac{1}{T}\int_0^T \mathrm{d}t\sqrt{\frac{\mathrm{Tr}(\rho_t^2H_t^2-(\rho_tH_t)^2)}{\mathrm{Tr}(\rho_t^2)}} ~~~\mathrm{and} \\
\Phi(\rho,\sigma)=\arccos\sqrt{\frac{\mathrm{Tr}(\rho\sigma)}{\mathrm{Tr}(\rho^2)}}\label{better bound}
\end{align}
We work with these different quantum speed limits for mixed quantum states and point out some examples where the newly derived quantum speed limit bound for mixed quantum states here performs better than the above bounds.

\subsection{Stronger Uncertainty Relations for general mixed quantum states}\label{SURGMS}

Robertson gave a rigorous and quantitative formulation of the heuristic Heisenberg's uncertainty principle, which are called the preparation uncertainty relations \cite{Robertson1929}. This is stated as the following. For any two noncommuting operators A and B, the Robertson-Schroedinger uncertainty relation for the state of the system $|\Psi\rangle$ is given by the following inequality:
\begin{align}
\Delta A^2\Delta B^2\geq |\frac{1}{2}\langle[A,B]\rangle|^2+|\frac{1}{2}\langle\{A,B\}\rangle-\langle A\rangle\langle B\rangle|^2,
\end{align}
where the averages and the variances are defined over the state of the quantum system $\rho$. However, this
uncertainty bound is not optimal. There have been several attempts to improve the bound. Here, we state a stronger bound obtained from an alternative uncertainty relation also called the Maccone-Pati uncertainty relation \cite{Pati2014} and is also state dependent. 
\begin{align}
\Delta A\Delta B\geq \frac{i}{2}\frac{\mathrm{Tr}(\rho[A,B])}{(1-\frac{1}{2}|\mathrm{Tr}(\rho^{\frac{1}{2}}(\frac{A}{\Delta A}\pm i \frac{B}{\Delta B})\sigma) |^2)},
\end{align}
where $\mathrm{Tr}(\rho^{\frac{1}{2}}\sigma)=0$ and $||\sigma||_2=1$. This uncertainty relation has been proved to be stronger than Robertson-Schrodinger uncertainty relation. It is optimized to an equality when maximized over all possible $\sigma$ possible, such that we have the optimized bound as 
\begin{align}
\Delta A\Delta B\geq \max_{\sigma}\frac{i}{2}\frac{\mathrm{Tr}(\rho[A,B])}{(1-\frac{1}{2}|\mathrm{Tr}(\rho^{\frac{1}{2}}(\frac{A}{\Delta A}\pm i \frac{B}{\Delta B})\sigma) |^2)}.
\end{align}
We can take the absolute values on both sides and then perform optimization, so that we get the following uncertainty relation
\begin{align}
\Delta A\Delta B\geq \max_{\sigma}\frac{1}{2}\frac{|\mathrm{Tr}(\rho[A,B])|}{|(1-\frac{1}{2}|\mathrm{Tr}(\rho^{\frac{1}{2}}(\frac{A}{\Delta A}\pm i \frac{B}{\Delta B})\sigma) |^2)|}.
\end{align}
We will use the above stronger uncertainty relations for mixed quantum states to derive a stronger version of quantum speed limits for mixed quantum states. See \cite{SURM} for the proof of the stronger uncertainty relations for mixed quantum states.

\section{Result: Stronger Quantum Speed Limit for unitarily driven mixed quantum states}

\theorem{The time evolution of a general mixed quantum state governed by a unitary operation generated by a Hamiltonian is given by the following equation
\begin{align}
\label{eqn:meq}
\tau\geq\tau_{SQSLM}= \frac{\sqrt{\mathrm{Tr}(\rho_0^2)}}{2\Delta H }\times\\ \nonumber \int_{s_0(0)}^{s_0(\tau)}\frac{\sin s_0(t)}{(1-R(t))\cos \frac{s_0(t)}{2}\sqrt{(1-\mathrm{Tr}(\rho_0^2)\cos^2 \frac{s_0(t)}{2})}}ds_0,
\end{align}
where $\tau_{SQSLM}$ stands as a short form for the stronger quantum speed limit for mixed quantum states and we have the following definitions of the quantities expressed in the above equation
\begin{align}\nonumber
s_0(t)=2\cos^{-1}|\sqrt{\frac{\mathrm{Tr}(\rho(0)\rho(t))}{\mathrm{Tr}(\rho_0^2)}}|,\\ \nonumber
\Delta H=\mathrm{Tr}(H^2\rho)-(\mathrm{Tr}(H\rho))^2\\ \nonumber 
R(t)=\frac{1}{2}|\mathrm{Tr}(\rho^{\frac{1}{2}}(\frac{A}{\Delta A}\pm i \frac{B}{\Delta B})\sigma) |^2, \\ \nonumber \mathrm{where} ~\mathrm{Tr}(\rho^{\frac{1}{2}}\sigma)=0 ~\mathrm{and}~||\sigma||_2=1,
\end{align}
where we have denoted $\rho_0=\rho(0)$, $\rho=\rho(t)$ and used this interchangeably everywhere, $||\sigma||_2=(\sum_{n\in I}\langle e_n|\sigma\sigma^\dagger|e_n\rangle)^{\frac{1}{2}}$, $\{|e_n\rangle\}$ forming a complete orthonormal basis in Hilbert space $\mathcal{H}$, $\sigma\in L^2(\mathcal{H})$, i.e., $\sigma$ belongs to the set of all Hilbert Schmidt linear operators.}
\proof {The proof of the above theorem goes as follows. We start by writing out the stronger uncertainty relation for mixed quantum states as is given by the following 
\begin{align}
\Delta A\Delta B\geq \frac{1}{2}\frac{|\mathrm{Tr}(\rho[A,B])|}{|(1-\frac{1}{2}|\mathrm{Tr}(\rho^{\frac{1}{2}}(\frac{A}{\Delta A}\pm i \frac{B}{\Delta B})\sigma) |^2)|},
\end{align}
See \cite{SURM} for the derivation of the above inequality. From the stronger uncertainty relation for mixed quantum states, we get the following
\begin{align}
\Delta A \Delta H (1-R(t))\geq \frac{1}{2}|\mathrm{Tr}(\rho[A,H])|, 
\end{align}
where we have defined $R(t)$ as the following
\begin{align}
R(t)=\frac{1}{2}|\mathrm{Tr}(\rho^{\frac{1}{2}}(\frac{A}{\Delta A}\pm i \frac{B}{\Delta B})\sigma) |^2
\end{align}
and have taken $A=\rho_0$ and $B=H$ for our purpose of deriving the stronger quantum speed limit for mixed quantum states. This particular choice of these operators help us to formulate our inequality for the quantum speed limit for mixed quantum states. Also for mixed quantum states, from Eahrenfest's theorem we get the following
\begin{align}
i\hbar\frac{d\mathrm{Tr}(\rho A)}{dt}=\mathrm{Tr}(\rho[A,H])
\end{align}
Therefore from the above equations, we get the following
\begin{align}
\Delta A \Delta H (1-R(t))\geq \frac{\hbar}{2}|\frac{d\langle A\rangle}{dt}|
\end{align}
The variance of the operator $A$ is then given by 
\begin{align}
\Delta A^2&=\mathrm{Tr}(\rho(0)^2\rho(t))-(\mathrm{Tr}(\rho(0)\rho(t)))^2\nonumber \\&=\mathrm{Tr}(\rho_0^2\rho_t)-(\mathrm{Tr}(\rho_0\rho_t))^2,
\end{align}
where we have used the notation $\rho(0)=\rho_0$ and $\rho(t)=\rho_t$. We can now take the following parametrization
\begin{align}
\langle A\rangle=\mathrm{Tr}(\rho(0)\rho(t))=\mathrm{Tr}(\rho_0^2)\cos^2\frac{s_0(t)}{2}.\label{expecA}
\end{align}
Now, using the equation of motion for the average of $A$
\begin{align}\nonumber
|\hbar \frac{d}{dt}\langle A\rangle|=|\langle[A,H]\rangle|,
\end{align}
where the averages are all with respect to the mixed quantum state $\rho$ and the quantum mechanical hermitian operator $A$ has no explicit time dependence. Thus, using Eq.\eqref{expecA}, we get
\begin{align}
|\frac{d\langle A\rangle}{dt}| =\mathrm{Tr}(\rho_0^2)\frac{\sin s_0(t)}{2}\frac{ds_0}{dt}
\end{align}
Now let us analyze the structure of $\Delta A^2$ as follows
\begin{align}
\Delta A^2=\mathrm{Tr}(\rho_0^2\rho_t)-(\mathrm{Tr}(\rho_0\rho_t))^2.
\end{align}
Let $\{|k\rangle\}$ be the eigenbasis from the singular value decomposition of the density matrix $\rho_0$. Then we have the following expression
\begin{align}
\rho_0=\sum_{k}\lambda_{k}|k\rangle\langle k| ~\mathrm{and}~
\rho_0^2=\sum_{k}\lambda_{k}^2|k\rangle\langle k|.
\end{align}
Using the above equation we obtain the following quantities
\begin{align}
\mathrm{Tr}(\rho_0\rho_t)&=\sum_{k}\lambda_{k}\langle k|\rho_t|k\rangle ~\mathrm{and}
\mathrm{Tr}(\rho_0^2\rho_t)&=\sum_{k}\lambda_{k}^2\langle k|\rho_t|k\rangle.
\end{align}
Since, we know that $0\leq\lambda_{k}^2\leq \lambda_{k}\leq 1~\forall ~k$ and also $\langle k|\rho_t|k\rangle\geq 0 ~\forall~k$ because $\rho_t$ is a positive operator. Therefore, we get the following inequality
\begin{align}
\mathrm{Tr}(\rho_0\rho_t)\geq\mathrm{Tr}(\rho_0^2\rho_t).
\end{align}
Adding $-(\mathrm{Tr}(\rho_0\rho_t))^2$ on both side of the above equation we get 
\begin{align}
\mathrm{Tr}(\rho_0\rho_t)-(\mathrm{Tr}(\rho_0\rho_t))^2&\geq 
\mathrm{Tr}(\rho_0^2\rho_t)-(\mathrm{Tr}(\rho_0\rho_t))^2\nonumber\\
&=\Delta A^2.
\end{align}
Now, using Eq.\eqref{expecA} we get
\begin{align}
\mathrm{Tr}(\rho_0^2)\cos^2 \frac{s_0(t)}{2}(1-\mathrm{Tr}(\rho_0^2)\cos^2 \frac{s_0(t)}{2})\geq\Delta A^2
\end{align}
Taking square root on both sides and multiplying by $\Delta H$ we get 
\begin{align}
\sqrt{\mathrm{Tr}(\rho_0^2)}\cos \frac{s_0(t)}{2}\sqrt{(1-\mathrm{Tr}(\rho_0^2)\cos^2 \frac{s_0(t)}{2})}\Delta H\geq \Delta A\Delta H.
\end{align}
From here, we get the following 
\begin{align}\nonumber 
\sqrt{\mathrm{Tr}(\rho_0^2)}\cos \frac{s_0(t)}{2}\sqrt{(1-\mathrm{Tr}(\rho_0^2)\cos^2 \frac{s_0(t)}{2})}\Delta H (1-R(t))\\ \nonumber \geq \Delta A\Delta H (1-R(t)),
\end{align}
since $(1-R(t))$ is a positive quantity here.
From the previous equations we get the following
\begin{align}\nonumber 
\sqrt{\mathrm{Tr}(\rho_0^2)}\cos \frac{s_0(t)}{2}\sqrt{(1-\mathrm{Tr}(\rho_0^2)\cos^2 \frac{s_0(t)}{2})}\Delta H (1-R(t))\\ \nonumber \geq \Delta A\Delta H (1-R(t))\geq \frac{\hbar}{2}|\frac{d\langle A\rangle}{dt}|=\mathrm{Tr}(\rho_0^2)\frac{\sin s_0(t)}{2}\frac{ds_0}{dt},
\end{align}
Therefore, from the above equations we get the following 
\begin{align}\nonumber 
\cos \frac{s_0(t)}{2}\sqrt{(1-\mathrm{Tr}(\rho_0^2)\cos^2 \frac{s_0(t)}{2})}\Delta H \geq \\ \nonumber \frac{\sqrt{\mathrm{Tr}(\rho_0^2)}}{(1-R(t))}\frac{\sin s_0(t)}{2}\frac{ds_0}{dt},
\end{align}
Integrating the above equation with respect to $t$ and $s$ over their corresponding regions on both sides, we get for the case of time independent Hamiltonian the following expression for quantum speed limit
\begin{align}\nonumber 
\tau\geq \frac{\sqrt{\mathrm{Tr}(\rho_0^2)}}{2\Delta H }\times\\ \nonumber \int_{s_0(0)}^{s_0(\tau)}\frac{\sin s_0(t)}{(1-R(t))\cos \frac{s_0(t)}{2}\sqrt{(1-\mathrm{Tr}(\rho_0^2)\cos^2 \frac{s_0(t)}{2})}}ds_0,
\end{align}
where the definitions of the parametrizations have been stated in the statement of the theorem. One can also derive the quantum speed limit bound for mixed quantum states in a different way. Writing out the previous equations and rearranging terms on the right hand side and the left hand side in a different way, it can be shown that the quantum speed limit bound for the mixed quantum states can also be written following the procedure as stated below step by step. We start from the following inequality after rearranging the terms
\begin{align}\nonumber 
\cos \frac{s_0(t)}{2}\sqrt{(1-\mathrm{Tr}(\rho_0^2)\cos^2 \frac{s_0(t)}{2})}\Delta H (1-R(t)) \geq \\ \nonumber \frac{\sqrt{\mathrm{Tr}(\rho_0^2)}\sin s_0(t)}{2}\frac{ds_0}{dt},
\end{align}
Integrating the above equation we get the following quantum speed limit bound for mixed quantum states
{\small{\begin{align}\nonumber 
\tau \geq \\ \nonumber \int_{s(0)}^{s(\tau)}\frac{\sqrt{\mathrm{Tr}(\rho_0^2)}\sin s_0(t)}{2\Delta H \cos \frac{s_0(t)}{2}\sqrt{(1-\mathrm{Tr}(\rho_0^2)\cos^2 \frac{s_0(t)}{2})} }ds_0+  
\int_{0}^{\tau}R(t){dt},
\end{align}}}
From the above equations, we get the following 
\begin{align}
\tau\geq \big[\frac{2 \cos^{-1}(\sqrt{\mathrm{Tr}(\rho_0^2)} \cos\frac{s_0}{2})}{\Delta H}\big]_{s(0)}^{s(\tau)}+\int_{0}^{\tau}R(t){dt}
\end{align}
Putting the values, we get the following equation for time independent Hamiltonians
\begin{align}
\tau\geq \big[\frac{2 (\cos^{-1}(\sqrt{\mathrm{Tr}(\rho_0\rho_t)} )-\cos^{-1}(\sqrt{\mathrm{Tr}(\rho_0^2)} ))}{\Delta H}\big]\\ \nonumber +\int_{0}^{\tau}R(t){dt}
\end{align}
It is easy to see that the above bound reduces to that of the stronger quantum speed limit bound for pure states when we take $\mathrm{Tr}(\rho_0^2)=1$, which performs better than the MT bound for pure quantum states.
}

\subsection{Method to find $\sigma$, such that $\mathrm{Tr}(\rho^{\frac{1}{2}}\sigma)=0$}

For the purpose of calculating our bound, we need to find ways to derive the structure of $\sigma$ or identify the set of $\sigma$ such that the condition $\mathrm{Tr}(\rho^{\frac{1}{2}}\sigma)=0$ is satisfied. In the preceding paragraphs, we find out two different ways to do so and apply them to examples thereafter. 

\subsubsection{Method I: $\rho$ and $\sigma$ $\in$ orthogonal subspaces}

In this section we derive the method that can be useful to find $\sigma$ such that the condition $\mathrm{Tr}(\rho^{\frac{1}{2}}\sigma)=0$ holds. First let us state the properties of $\sigma$ that should be satisfied in that case. It should satisfy $||\sigma||_2=1$, where $ ||\sigma||_2=(\sum_{n\in I}\langle e_n|\sigma^\dagger\sigma|e_n\rangle)^{\frac{1}{2}}$ and $\sigma\in L^2(H)$. Let us take the following definitions
\begin{align}
\rho=\sum_{k}\lambda_k|k\rangle\langle k|, ~~|' ~~\rho^{\frac{1}{2}}=\sum_{k}\lambda_k^{\frac{1}{2}}|k\rangle\langle k|,
\end{align}
where we have $\sum_k\lambda_k=1$ fixed by the normalization constraint of $\rho$ and we have taken the positive square root of $\lambda_k$. Note that we have written $\rho$ in its eigenbasis and can be reverted back to any other basis by unitary transformation and the same holds for $\rho^{\frac{1}{2}}$ in a corresponding way. In this way $\rho^{\frac{1}{2}}$ is also a positive semidefinite Hermitian operator as $\rho$. Let us denote $\lambda_k^{\frac{1}{2}}=\eta_k$ for convenience. Therefore, following this notation, we have 
\begin{align}
\rho^{\frac{1}{2}}=\sum_{k}\eta_k|k\rangle\langle k|.
\end{align}
Therefore from the condition $\mathrm{Tr}(\rho^{\frac{1}{2}}\sigma)=0$, we get 
\begin{align}
\mathrm{Tr}(\sum_k\eta_k|k\rangle\langle k|\sigma)=0.
\end{align}
This translates to the following condition
\begin{align}
\sum_k\eta_k\langle k|\sigma|k\rangle=0.
\end{align}
We know that $\eta_k\geq 0~\forall~ k$ from our own constraint which we have specifically chosen that we only take the positive square root of $\lambda_k ~\forall ~ k$ as $\eta_k$.
Also when we impose the condition that $\sigma$ is also a positive operator, then we get the condition that $\langle k|\sigma|k\rangle\geq 0~\forall~k$. One of the ways this condition can be obtained is that if $\rho$ and $\sigma$ are chosen from orthogonal subspaces. Let us note here that $\rho$ is fixed here and we do not have a choice to fix $\rho$ and we only have the freedom to choose any $\sigma$ from the orthogonal subspace to that of $\rho$. As a result we can optimize our bound for the stronger quantum speed limit over all possible choices of such $\sigma$ chosen from the orthogonal subspaces to that of $\rho$. For mixed quantum states, this choice of $\sigma$ becomes relevant only in higher dimensional Hilbert spaces than the qubit space.

\subsubsection{Method II: A form of $\sigma$ written directly in terms of $\rho$ and Hermitian operators.}

There is another method that allows one to derive an operator that satisfies the condition $\mathrm{Tr}(\rho^{\frac{1}{2}}\sigma)=0$ in a more easier way. This set of $\sigma$ can be written down in the following form 
\begin{align}
\sigma=\frac{O-\langle O\rangle}{\Delta O}\rho^{\frac{1}{2}},
\end{align}
where, $O$ is any Hermitian operator. This way the conditions $\mathrm{Tr}(\rho^{\frac{1}{2}}\sigma)=0$ and $\mathrm{Tr}(\sigma\sigma^\dagger)=1$ are satisfied automatically. The proof of this claim in given in the following paragraph.
\proof{The proof of the first condition $\mathrm{Tr}(\rho^{\frac{1}{2}}\sigma)=0$ goes as follows. 
\begin{align}\nonumber 
\mathrm{Tr}(\rho^{\frac{1}{2}}\sigma)=\mathrm{Tr}(\rho^{\frac{1}{2}}\frac{O-\langle O\rangle}{\Delta O}\rho^{\frac{1}{2}}) 
=\frac{1}{\Delta O}\mathrm{Tr}(\rho(O-\langle O\rangle))=0
\end{align}
Now we show that the $\sigma$ defined in this way also satisfies the condition $\mathrm{Tr}(\sigma\sigma^\dagger)=1$. This is as follows.
\begin{align}\nonumber
\mathrm{Tr}(\sigma\sigma^\dagger)=\mathrm{Tr}(\frac{O-\langle O\rangle}{\Delta O}\rho^{\frac{1}{2}}(\frac{O-\langle O\rangle}{\Delta O}\rho^{\frac{1}{2}})^\dagger)\\ \nonumber
=\mathrm{Tr}((\frac{O-\langle O\rangle}{\Delta O})\rho^{\frac{1}{2}}\rho^{\frac{1}{2}}(\frac{O-\langle O\rangle}{\Delta O}))\\ \nonumber
=\mathrm{Tr}((\frac{O-\langle O\rangle}{\Delta O})\rho(\frac{O-\langle O\rangle}{\Delta O}))
=\mathrm{Tr}(\rho(\frac{O-\langle O\rangle}{\Delta O})^2)=1
\end{align}
}
As a result, we have derived another set of operators $\sigma$ that satisfies the required conditions essential for deriving the stronger quantum speed limit bound for mixed quantum states. Also we see that since $O$ can be any Hermitian operator, therefore we can have a large set of $\sigma$ as stated above that satisfies our required criterion based on the different Hermitian operators that we can choose. Using this way of finding $\sigma$, the stronger quantum speed limit bound is simplified further as follows. We start with the expression of $R(t)$ which is as follows 
\begin{align}
R(t)=\frac{1}{2}|\mathrm{Tr}(\rho^{\frac{1}{2}}(\frac{A}{\Delta A}\pm i \frac{B}{\Delta B})\sigma) |^2.
\end{align}
We put the expression of $\sigma$ as described in this section and find the following expression for $R(t)$
\begin{align}
R(t)=\frac{1}{2}|\mathrm{Tr}(\rho^{\frac{1}{2}}(\frac{A}{\Delta A}\pm i \frac{B}{\Delta B})(\frac{O-\langle O\rangle}{\Delta O}\rho^{\frac{1}{2}})) |^2.
\end{align}
Using the cyclic property of the trace function, therefore we arrive at the following simplified version of $R(t)$
\begin{align}
R(t)=\frac{1}{2}|\mathrm{Tr}(\rho(\frac{A}{\Delta A}\pm i \frac{B}{\Delta B})(\frac{O-\langle O\rangle}{\Delta O})) |^2.
\end{align}
The above expression is clearly computationally much more efficient and less time consuming, where for the calculation of the stronger speed limit bound for mixed quantum states, one does not have to compute the square root of $\rho$, making the calculation of the bound more efficient, fast and simple. We will apply this technique for the examples in the next section.

\section{Examples}

\subsection{Random Hamiltonians }\label{examples}

In this section, we calculate and compare the bound given by the tighter quantum speed limit bound with that of the MT like bound of mixed state generalization using random Hamiltonians from the Gaussian Unitary Ensemble or GUE in short. Random Hamiltonians from GUE have found use in many different areas. But our reason for choosing Hamiltonians randomly from GUE is that they give vaild Hamiltonians that are also diverse such that we can show the performance of our stronger quantum speed limit bound for mixed quantum states and unitary evolutions for diverse cases. 

Mathematically, a random Hamiltonian is a $\mathrm{D}\times \mathrm{D}$ Hermitian operator $H$ in $\mathrm{D}\times \mathrm{D}$ dimensional Hilbert space, drawn from a Gaussian unitary ensemble (GUE). The GUE is described by the following probability distribution function 
 \begin{align}
 P(H)= Ce^{-\frac{D}{2}\mathrm{Tr}(H^2)}
 \end{align}
 where $C$ is the normalization constant and the elements of $H$ are drawn from the Gaussian probability distribution. In this way $H$ is also Hermitian. A random Hamiltonian dynamics is an unitary time- evolution generated by a fixed time-independent GUE Hamiltonian. 
 
\begin{figure*}
\begin{subfigure}{0.485\textwidth}
\includegraphics[width=\linewidth]{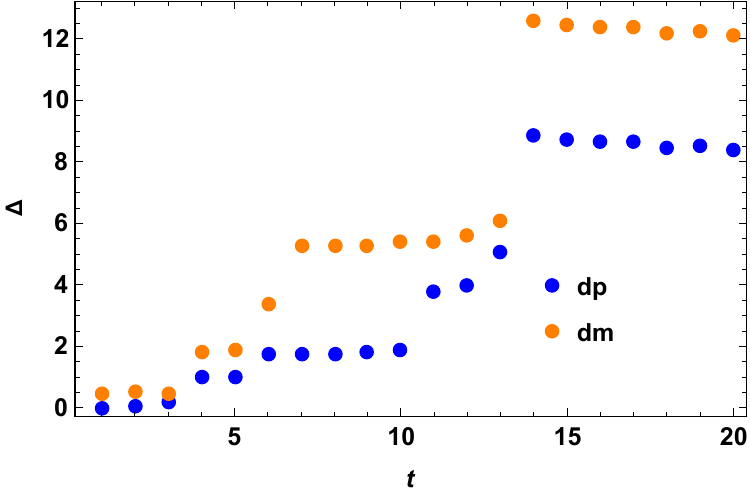}
\caption{Random Hamiltonian example 1 with a fixed initial mixed quantum state over full time range with no optimization over the Hermitian matrices $O$. The vertical axis represents $\Delta=\tau_{SQSL}-\tau_{MTL}$, for the case of both blue (when $R(t)$ in $\ref{eqn:meq}$ has $+$ sign inside) and orange (when $R(t)$ in $\ref{eqn:meq}$ has $-$ sign inside) data points. The horizontal axis $t$ represents evolution time.}
\label{fig:sfig1a}
\end{subfigure}
\hfill
\begin{subfigure}{0.485\textwidth}
\includegraphics[width=\linewidth]{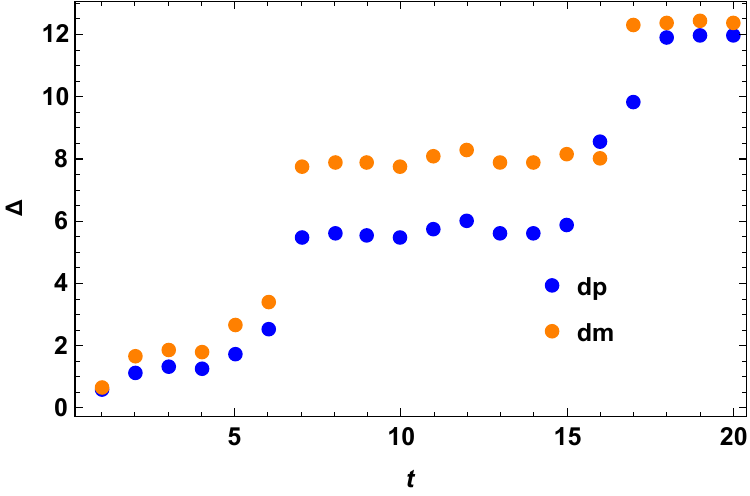}
\caption{Random Hamiltonian example 2 with a fixed initial mixed quantum state over full time range with optimization over three Hermitian matrices $O$. The vertical axis represents $\Delta=\tau_{SQSL}-\tau_{MTL}$, for the case of both blue (when $R(t)$ in $\ref{eqn:meq}$ has $+$ sign inside) and orange (when $R(t)$ in $\ref{eqn:meq}$ has $-$ sign inside) data points. The horizontal axis $t$ represents evolution time.}
\label{fig:sfig1b}
\end{subfigure}
\caption{Random Hamiltonian examples.}
\label{fig:fig1}
\end{figure*}

 We take the Hilbert space of dimension 3 for our numerical example as shown in Fig.\ref{fig:fig1}. The initial state is taken as the following 
 \begin{align}
 \rho_0=0.2|0\rangle\langle 0|+0.5|1\rangle\langle 1|+0.3|2\rangle\langle 2|
 \end{align}
 Following the second method of generating appropriate $\sigma$ using a set of Hermitian operators $O$, we obtain the quantum speed limit bound for the mixed quantum states. We compare the performance of our optimized bound with the previous bounds and non optimized version of our bound as given in the figures. From both the subfigures \ref{fig:sfig1a} and \ref{fig:sfig1b} in Fig.\ref{fig:fig1}, we clearly see that our theory is correct and we have $\Delta=\tau_{SQSL}-\tau_{MT}$ as always positive, showing that the stronger quantum speed limit bound always outperforms the MT like bound for mixed quantum states and unitary evolution. In Fig.\ref{fig:fig1}, at $t=0$, all the values of $\Delta$ are zero because all the random Hamiltonians start with being identity at $t=0$. All the Hamiltonians taken here are time independent by construction. In subfigure \ref{fig:sfig1b}, we perform an optimization over different sets of $\sigma$ so as to get a better bound, whereas in subfigure \ref{fig:sfig1a}, we still get good results even without any optimization. In the figures and everywhere later in the later examples in the next sections, $dp$ represents the difference of our bound with the MT like bound as in Eq.(\ref{eqn:mtl}) when one uses a $+$ sign in front of $R(t)$ and $dm$ represents the difference of our bound with the MT like bound as in Eq.(\ref{eqn:mtl}) when one uses a $-$ sign in front of $R(t)$, unless stated otherwise. We also perform optimization of our bound over small sets of $\sigma$ and note that our bound performs better with or without optimization in these cases, as exemplified by the figures. When we perform optimization, it is simple and easily completed within about a minute in most cases for such small sets of $\sigma$ such as $5$ or $10$ number of $\sigma$ as stated in the caption of the figures. This makes our method computationally practical and feasible. This simple optimization  also gives noticeable improvement on the bounds as demonstrated by the figures, in this example as well as other examples, in the following sections. However, since we cannot tell a priori which optimized version will give the best bound and in which region due to no closed form of the optimized version for arbitrary Hamiltonian, as a result we keep this as an open question for future investigation.
\begin{figure*}\label{fig2}
\begin{subfigure}{0.485\textwidth}
\includegraphics[width=\linewidth]{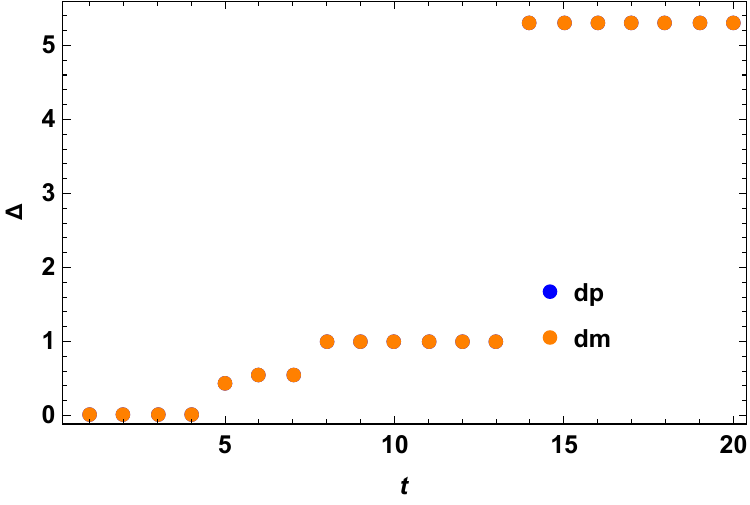}
\caption{Anisotropic Heisenberg spin chain example 1 with a fixed initial mixed quantum state over full time range with no optimization over the Hermitian matrices $O$. The vertical axis represents $\Delta=\tau_{SQSL}-\tau_{MTL}$, for the case of both blue (when $R(t)$ in $\ref{eqn:meq}$ has $+$ sign inside) and orange (when $R(t)$ in $\ref{eqn:meq}$ has $-$ sign inside) data points. The horizontal axis $t$ represents evolution time. Blue and orange dots coincide.}
\label{fig:sfig2a}
\end{subfigure}
\hfill
\begin{subfigure}{0.485\textwidth}
\includegraphics[width=\linewidth]{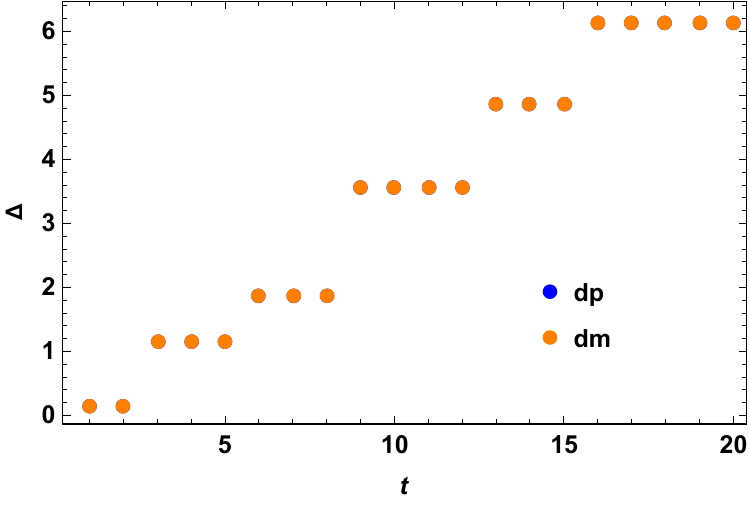}
\caption{Anisotropic Heisenberg spin chain example 1 with a fixed initial mixed quantum state over full time range with optimization over three Hermitian matrices $O$. The vertical axis represents $\Delta=\tau_{SQSL}-\tau_{MTL}$, for the case of both blue (when $R(t)$ in $\ref{eqn:meq}$ has $+$ sign inside) and orange (when $R(t)$ in $\ref{eqn:meq}$ has $-$ sign inside) data points. The horizontal axis $t$ represents evolution time. Blue and orange dots coincide.}
\label{fig:sfig2b}
\end{subfigure}
\caption{Anisotropic Heisenberg spin chain Example 1.}
\label{fig:fig2}
\end{figure*}
\begin{figure*}
\begin{subfigure}{0.485\textwidth}
\includegraphics[width=\linewidth]{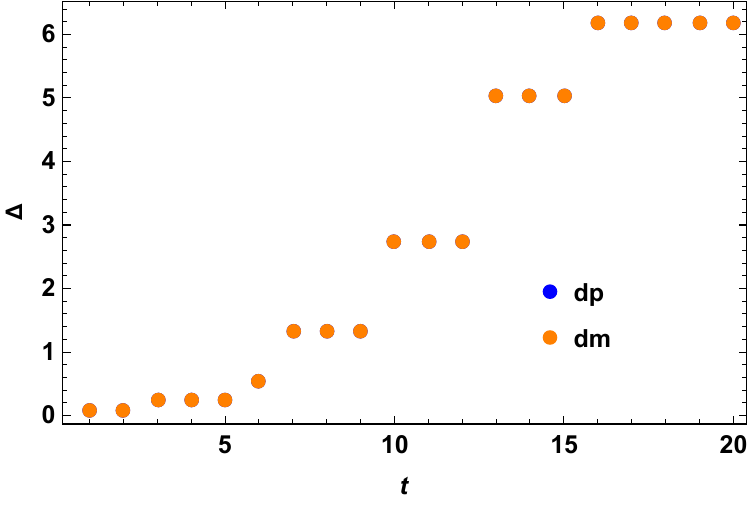}
\caption{Anisotropic Heisenberg spin chain example 2 with a fixed initial mixed quantum state over full time range with no optimization over the Hermitian matrices $O$. The vertical axis represents $\Delta=\tau_{SQSL}-\tau_{MTL}$, for the case of both blue (when $R(t)$ in $\ref{eqn:meq}$ has $+$ sign inside) and orange (when $R(t)$ in $\ref{eqn:meq}$ has $-$ sign inside) data points. The horizontal axis $t$ represents evolution time. Blue and orange dots coincide. }
\label{fig:sfig3a}
\end{subfigure}
\hfill
\begin{subfigure}{0.485\textwidth}
\includegraphics[width=\linewidth]{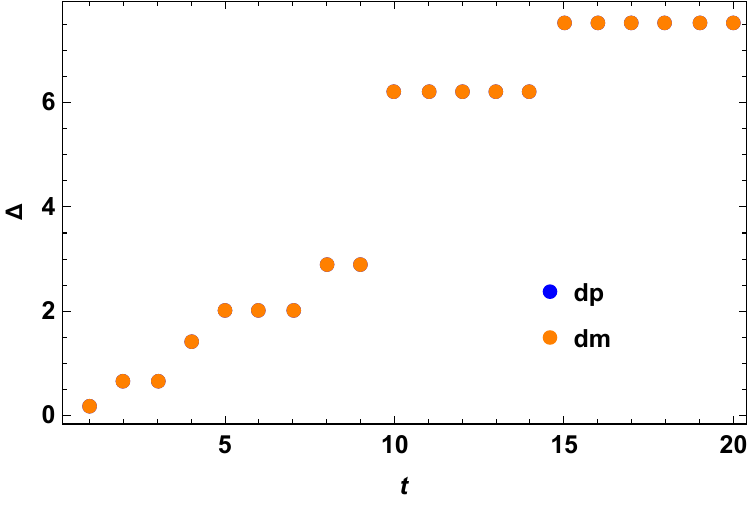}
\caption{Anisotropic Heisenberg spin chain example 2 with a fixed initial mixed quantum state over full time range with optimization over three Hermitian matrices $O$. The vertical axis represents $\Delta=\tau_{SQSL}-\tau_{MTL}$, for the case of both blue (when $R(t)$ in $\ref{eqn:meq}$ has $+$ sign inside) and orange (when $R(t)$ in $\ref{eqn:meq}$ has $-$ sign inside) data points. The horizontal axis $t$ represents evolution time. Blue and orange dots coincide. }
\label{fig:sfig3b}
\end{subfigure}
\caption{Anisotropic Heisenberg spin chain Example 2.}
\label{fig:fig3}
\end{figure*}
 
\subsection{Anisotropic multiqubit Heisenberg spin chain }

A lot of attention has been devoted to the study of graph states, which play an important and central resource in quantum error correction, quantum cryptography and practical quantum metrology in the presence of noise. As a result, owing to its importance in quantum information processing tasks, we write here the entangling Hamiltonian of the graph state generation for the multiqubit case as follows. 
\begin{align}
H=\sum_{i=1}^N\lambda^z_i\sigma^z_i+\sum_{i=1}^N\lambda^{zz}\sigma^z_i\sigma^z_{i+1}-\\ \nonumber\sum_{i=1}^N\lambda^{xx}\sigma^x_i\sigma^x_{i+1}-\sum_{i=1}^N\lambda^{yy}\sigma^y_i\sigma^y_{i+1}
\end{align}
In terms of experiements, the above Hamiltonian is used in the physical implementation of optical lattice of ultracold bosonic atoms. This is also the anisotropic Heisenberg spin model in the optical lattice model which can be written down in appropriate way using the creation and the annihilation operators. The Hamiltonian has the local terms as well as the interaction terms and in general for $N$ spins which can be mapped to $N$ qubits. In general, the coefficients $\{\lambda\}$ are time dependent. However for simplicity we take this to be time independent in our case and calculate the quantum speed limit bound for evolution under this Hamiltonian for initially mixed quantum states.
 
We take the Hilbert space of dimension 4 for numerical example 1 as shown in the subfigures \ref{fig:sfig2a} and \ref{fig:sfig2b} of Fig.\ref{fig:fig2}, i.e., for the case of two qubits. The initial state is taken as the following 
\begin{align}
\rho_0=0.7|0\rangle\langle 0|+0.1|1\rangle\langle 1|+0.1|2\rangle\langle 2|+0.1|3\rangle\langle 3|
\end{align}
 Following the second method of generating appropriate $\sigma$, we obtaining the quantum speed limit bound for the mixed quantum states. We check our bound for  initial mixed quantum state as above under the action of the anisotropic Heisenberg spin chain Hamiltonian and compare the performance of our optimized bound with the previous bound. From the figures as in \ref{fig:sfig2a} and \ref{fig:sfig2b} of Fig.\ref{fig:fig2}, we clearly see that our theory is correct and we have $\Delta=\tau_{SQSL}-\tau_{MTL}$ as always positive, showing that the tighter quantum speed limit bound always outperforms the MT like bound for mixed quantum states.  The same holds for the example 2 as given in \ref{fig:sfig3a} and \ref{fig:sfig3b} of Fig.\ref{fig:fig3}, where a different instance of the anisotrpic Heisenberg spin has been considered with a different set of parameters but with the same underlying model as stated here. Since we cannot tell a priori which optimized version will give the best bound and in which region, as a result we keep this as an open question for future investigation.
\begin{figure*}
\begin{subfigure}{0.485\textwidth}
\includegraphics[width=\linewidth]{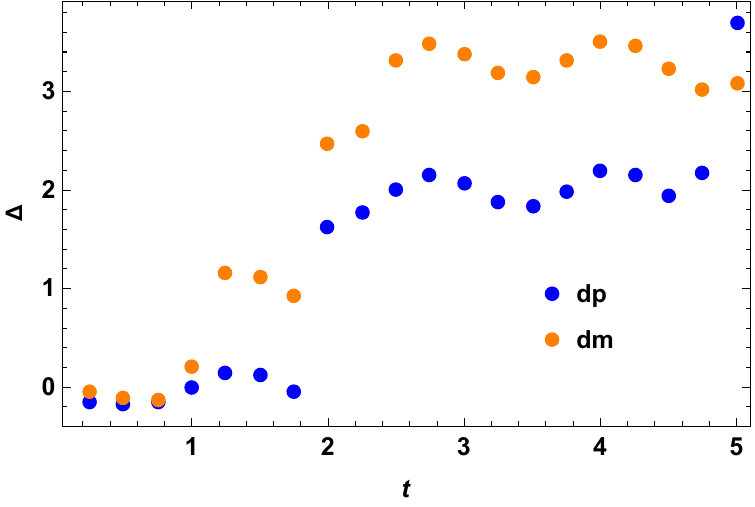}
\caption{Perfect state transfer Hamiltonian evolution of an initial mixed quantum states in two qubit Hilbert space with no optimization over random Hermitian operators $O$. The vertical axis represents $\Delta=\tau_{SQSL}-\tau_{MTL}$, for the case of both blue (when $R(t)$ in $\ref{eqn:meq}$ has $+$ sign inside) and orange (when $R(t)$ in $\ref{eqn:meq}$ has $-$ sign inside) data points. The horizontal axis $t$ represents evolution time.}
\label{fig:sfig4a}
\end{subfigure}
\hfill
\begin{subfigure}{0.485\textwidth}
\includegraphics[width=\linewidth]{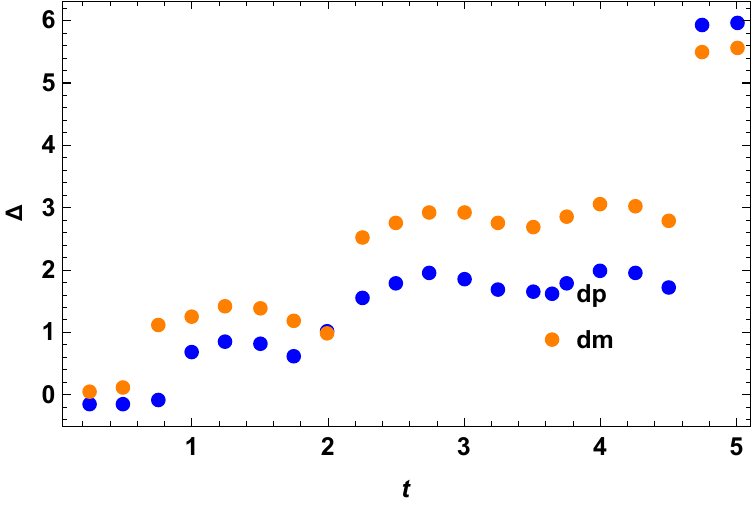}
\caption{Perfect state transfer Hamiltonian evolution of an initial mixed quantum states in two qubit Hilbert space with optimization over 3 random Hermitian operators $O$. The vertical axis represents $\Delta=\tau_{SQSL}-\tau_{MTL}$, for the case of both blue (when $R(t)$ in $\ref{eqn:meq}$ has $+$ sign inside) and orange (when $R(t)$ in $\ref{eqn:meq}$ has $-$ sign inside) data points. The horizontal axis $t$ represents evolution time.}
\label{fig:sfig4b}
\end{subfigure}
\caption{Perfect state transfer Hamiltonian example.}
\label{fig:fig4}
\end{figure*}

\begin{figure*}
\begin{subfigure}{0.495\textwidth}
\includegraphics[width=\linewidth]{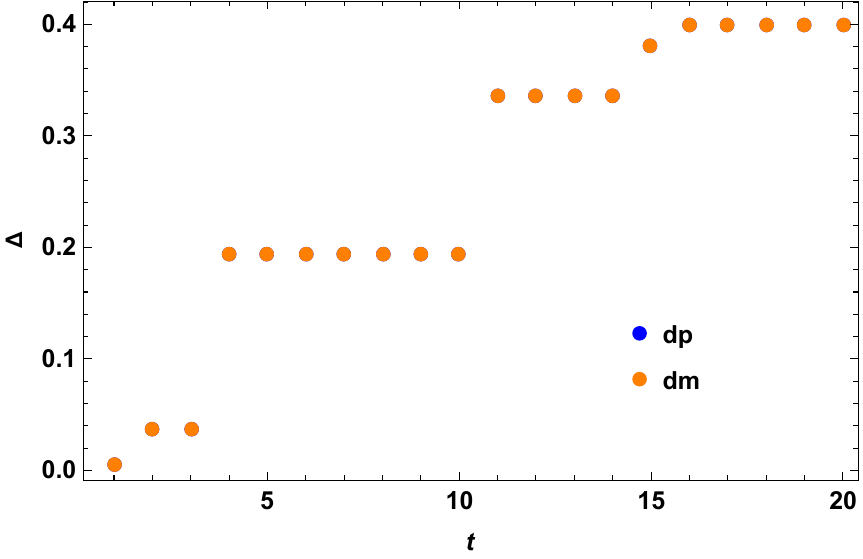}
\caption{Hamiltonian evolution of an initial separable mixed quantum states in two qutrit Hilbert space, according to Hamiltonian in Eq. (\ref{eqn:hes1}), without any optimization over random Hermitian operators $O$. The vertical axis represents $\Delta=\tau_{SQSL}-\tau_{MTL}$, for the case of both blue (when $R(t)$ in $\ref{eqn:meq}$ has $+$ sign inside) and orange (when $R(t)$ in $\ref{eqn:meq}$ has $-$ sign inside) data points. The horizontal axis $t$ represents evolution time. The blue and the orange dots coincide here. }
\label{fig:sfighea}
\end{subfigure}
\hfill
\begin{subfigure}{0.475\textwidth}
\includegraphics[width=\linewidth]{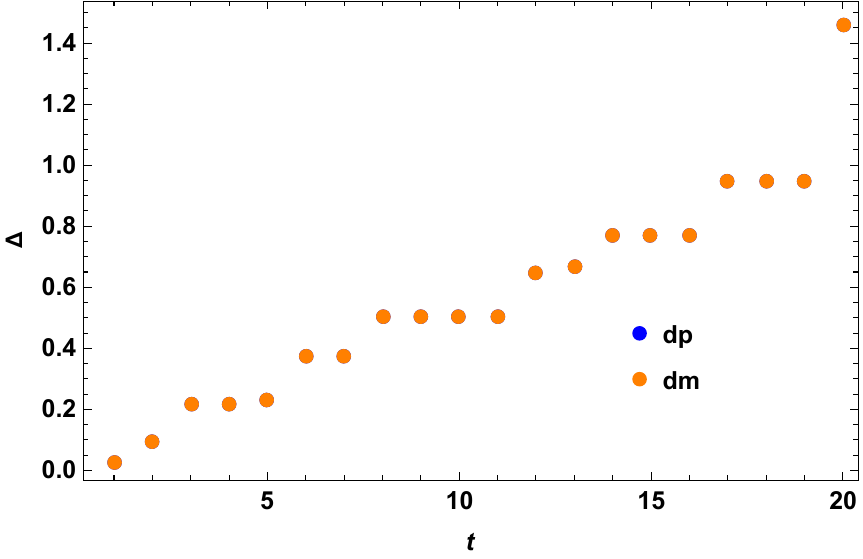}
\caption{Perfect state transfer Hamiltonian evolution of an initial mixed quantum states in two qubit Hilbert space, according to Hamiltonian in Eq. (\ref{eqn:hes1}), with optimization over 5 random Hermitian operators $O$. The vertical axis represents $\Delta=\tau_{SQSL}-\tau_{MTL}$, for the case of both blue (when $R(t)$ in $\ref{eqn:meq}$ has $+$ sign inside) and orange (when $R(t)$ in $\ref{eqn:meq}$ has $-$ sign inside) data points. The horizontal axis $t$ represents evolution time. The blue and the orange dots coincide here.}
\label{fig:sfigheb}
\end{subfigure}
\caption{Hamiltonian evolution of initial separable mixed quantum state according to Hamiltonian in Eq. (\ref{eqn:hes1}).}
\label{fig:fighe}
\end{figure*}

\begin{figure*}
\begin{subfigure}{0.485\textwidth}
\includegraphics[width=\linewidth]{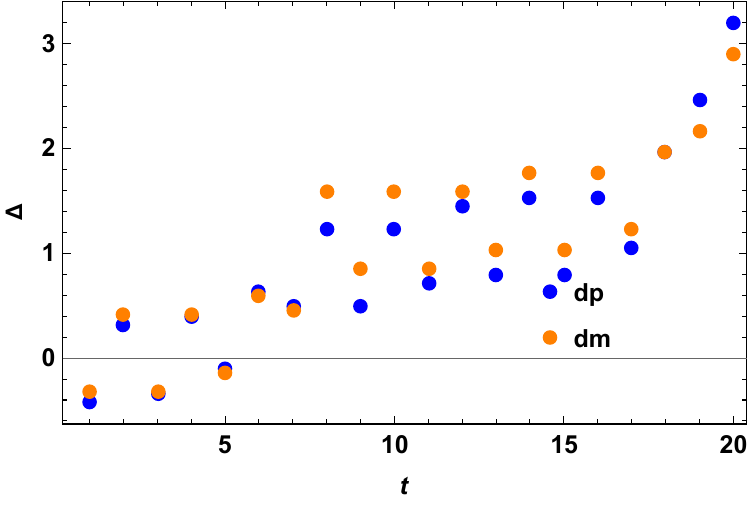}
\caption{Two qubit CNOT gate Hamiltonian evolution of an initial mixed quantum states in two qubit Hilbert space with no optimization over random Hermitian operators $O$. The vertical axis represents $\Delta=\tau_{SQSL}-\tau_{MTL}$, for the case of both blue (when $R(t)$ in $\ref{eqn:meq}$ has $+$ sign inside) and orange (when $R(t)$ in $\ref{eqn:meq}$ has $-$ sign inside) data points. The horizontal axis $t$ represents evolution time.}
\label{fig:sfig5a}
\end{subfigure}
\hfill
\begin{subfigure}{0.485\textwidth}
\includegraphics[width=\linewidth]{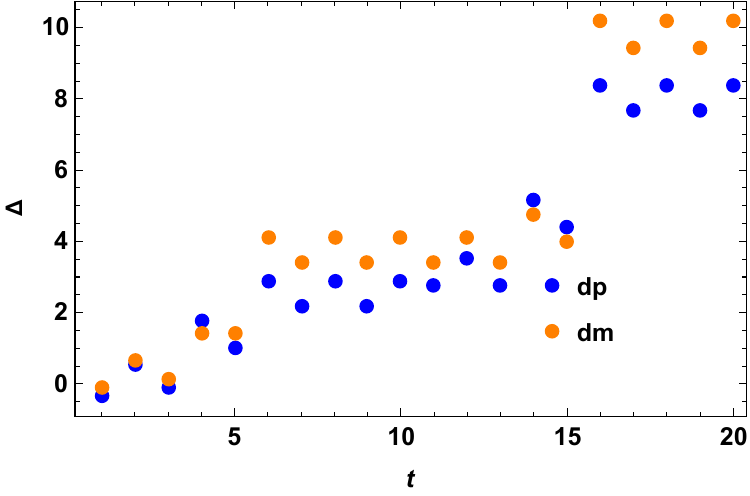}
\caption{Two qubit CNOT gate Hamiltonian evolution of an initial mixed quantum states in two qubit Hilbert space with optimization over 5 random Hermitian operators $O$. The vertical axis represents $\Delta=\tau_{SQSL}-\tau_{MTL}$, for the case of both blue (when $R(t)$ in $\ref{eqn:meq}$ has $+$ sign inside) and orange (when $R(t)$ in $\ref{eqn:meq}$ has $-$ sign inside) data points. The horizontal axis $t$ represents evolution time.}
\label{fig:sfig5b}
\end{subfigure}
\caption{Two qubit CNOT gate Hamiltonian example.}
\label{fig:fig5}
\end{figure*}
\begin{figure*}
\begin{subfigure}{0.485\textwidth}
\includegraphics[width=\linewidth]{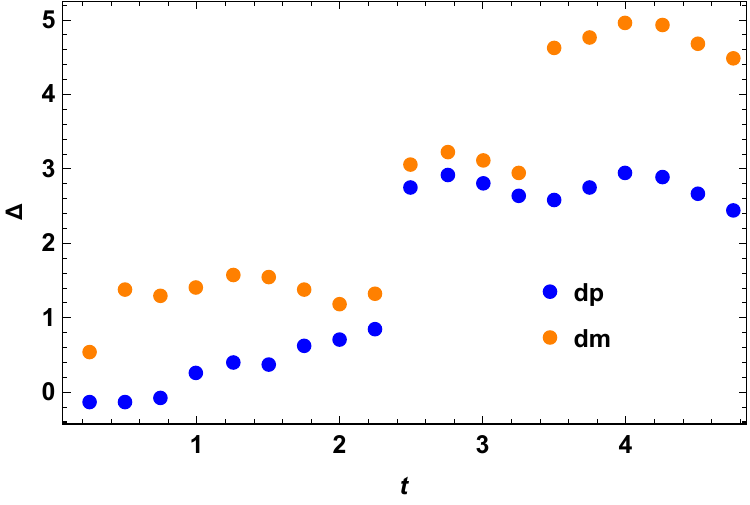}
\caption{Difference with the second existing quantum speed limit bound with our stronger quantum speed limit bound for the perfect state transfer Hamiltonian with optimization over 5 random Hermitian operators $O$. The vertical axis represents $\Delta=\tau_{SQSL}-\tau_{PRE1}$, where $\tau_{PRE1}$ is given by Eq.(\ref{eqn:pre1}), for the case of both blue (when $R(t)$ in $\ref{eqn:meq}$ has $+$ sign inside) and orange (when $R(t)$ in $\ref{eqn:meq}$ has $-$ sign inside) data points. The horizontal axis $t$ represents evolution time. It looks very similar to the next one, but there is a small difference which is shown in the next plot. }
\label{fig:sfig6a}
\end{subfigure}
\hfill
\begin{subfigure}{0.485\textwidth}
\includegraphics[width=\linewidth]{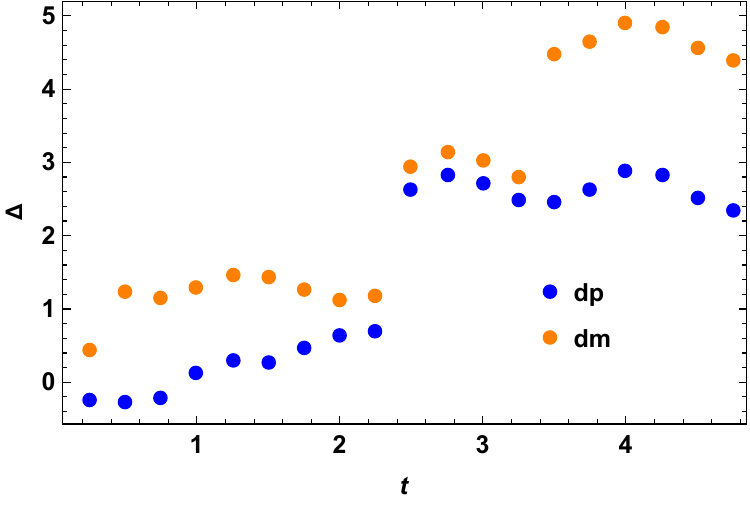}
\caption{Difference with the third existing quantum speed limit bound with our stronger quantum speed limit bound for the perfect state transfer Hamiltonian with optimization over 5 random Hermitian operators $O$. The vertical axis represents $\Delta=\tau_{SQSL}-\tau_{PRE2}$, where $\tau_{PRE2}$ is given by Eq.(\ref{eqn:pre2}), for  the case of both blue (when $R(t)$ in $\ref{eqn:meq}$ has $+$ sign inside) and orange (when $R(t)$ in $\ref{eqn:meq}$ has $-$ sign inside) data points. The horizontal axis $t$ represents evolution time. It looks very similar to the previous one, but there is a small difference which is shown in the next plot.}
\label{fig:sfig6b}
\end{subfigure}
\caption{Difference with the third existing quantum speed limit bound with our stronger quantum speed limit bound for the perfect state transfer Hamiltonian.}
\label{fig:fig6}
\end{figure*}
\begin{figure*}
\begin{subfigure}{0.495\textwidth}
\includegraphics[width=\linewidth]{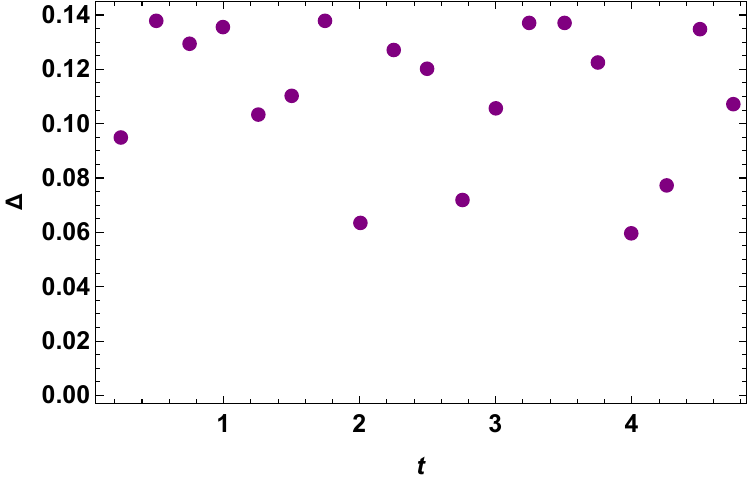}
\caption{Difference between the `difference between second existing quantum speed limit bound with our stronger quantum speed limit bound and the difference between third existing quantum speed limit bound with our stronger quantum speed limit bound' for the perfect state transfer Hamiltonian for minus sign in $R(t)$. The vertical axis represents $\Delta=(\Delta_{SQSL}-\Delta_{PRE1})-(\Delta_{SQSL}-\Delta_{PRE2})=(\Delta_{PRE2}-\Delta_{PRE1})$. The horizontal axis represents the evolution time.  }
\label{fig:sfig7a}
\end{subfigure}
\hfill
\begin{subfigure}{0.475\textwidth}
\includegraphics[width=\linewidth]{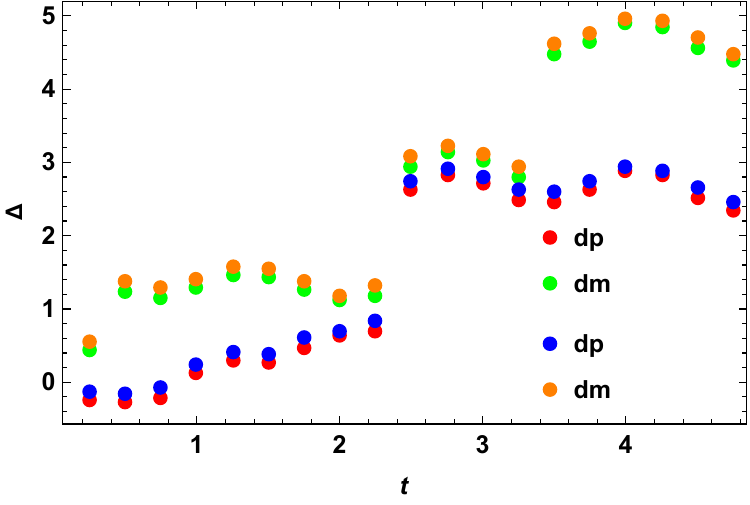}
\caption{Difference between the stronger speed limit bound using plus sign in $R(t)$ and the stronger speed limit bound using minus sign in $R(t)$ for mixed quantum states for the perfect state transfer Hamiltonian from the second and the third previous quantum speed limit bounds. The vertical axis represents $\Delta=(\Delta_{SQSL}-\Delta_{PRE1})$or$ (\Delta_{SQSL}-\Delta_{PRE2})$ as represented by different colours explained in section \ref{comparison}. The horizontal axis represents the evolution time. }
\label{fig:sfig7b}
\end{subfigure}
\caption{Difference with the stronger speed limit bound for plus sign in $R(t)$ with the stronger speed limit bound for minus sign in $R(t)$ for mixed quantum states for the perfect state transfer Hamiltonian from the second and the third previous quantum speed limit bounds.}
\label{fig:fig7}
\end{figure*}
 
\subsection{Perfect state transfer Hamiltonian}

Here, we take the example of a Hamiltonian which is useful for the case of perfect quantum state transfer, as quantum state transfer is one of the important quantum information processing tasks. The Hamiltonian describing the case of perfect state transfer is given by the following 
\begin{align}
\label{eqn:pst1}
H=\sum_{n=1}^{N-1} J_n\sigma_n^z\sigma^z_{n+1}+\sum_{n=1}^{N} B_n\sigma_n^x,
\end{align}
where $N$ is the number of qubits.
As specific numerical examples, we take the Hilbert space of dimension 4, i.e., for the case of two qubits. In this case, we take $J_k=\frac{1}{2}, B_k=\frac{1}{2}$ and then the Hamiltonian reads as the following for the case of two qubits as 
 \begin{align}
 \label{eqn:pst2}
 H=J_1(\sigma^z\otimes\sigma^z)+B_1(\sigma^x\otimes \mathbb{I})+B_2(\mathbb{I}\otimes\sigma^x).
 \end{align}
The initial state is taken as the following
 \begin{align}
 \label{eqn:pst3}
 \rho_0=0.7|0\rangle\langle 0|+0.1|1\rangle\langle 1|+0.1|2\rangle\langle 2|+0.1|3\rangle\langle 3|.
 \end{align}
We obtain the quantum speed limit bound for the mixed quantum states in the similar procedure as the other examples stated before. We check our bound for  initial mixed quantum state as stated above under the action of the quantum walker Hamiltonian as stated before and compare the performance of our optimized bound with the previous MT like bound for mixed quantum states. From the subfigures \ref{fig:sfig4a} and \ref{fig:sfig4b} of Fig.\ref{fig:fig4}, we clearly see that our theory is correct and we have $\Delta=\tau_{SQSL}-\tau_{MTL}$ as always positive, showing that the tighter quantum speed limit bound always outperforms the MT like (MTL) bound for mixed quantum states. 

\subsection{Hamiltonian evolution of a separable state}

Here, we take the example of another type of Hamiltonian  which drives the evolution of an initially mixed quantum state which we take to be a separable quantum state. The Hamiltonian describing this case is given by the following 
\begin{align}
\label{eqn:hes1}
H=\sum_{i=1}^{M} H_i ~~~, ~~~H_i=\omega\hbar \sum_{n=0}^{N-1} n |n\rangle\langle n|
\end{align}
where $M$ is the number of qubits and $N$ is the dimension of each subsystem.
As we mentioned, we take the initial state as a separable mixed state. This choice bears no particular importance. For our case of numerical example, we take the case of a quantum system of two qutrits. Even for this case of two qutrits, the derivation of the stronger quantum speed limit for mixed states is done within a fraction of a minute, even for an optimization over a set of 5 number of $\sigma$ operators. This implies that the derivation of the quantum speed limit for mixed quantum states can be done for a wide variety of quantum systems of different dimensions, in this case the dimension being 9. We demonstrate here a particular example by taking the following initial quantum state
 \begin{align}
 \label{eqn:hes3}
 \rho_0=a|0\rangle\langle 0|+b|1\rangle\langle 1|+c|2\rangle\langle 2|\\ +(1-a-b-c-d-e)|3\rangle\langle 3|+d|7\rangle\langle 7|+e|8\rangle\langle 8|,
  \end{align}
where we have the following parameters $a = 0.175, ~b = 0.25, ~c = 0.15, ~d = 0.105,~e = 0.255$. We have also set $\omega\hbar=1$ without any loss of generality. The choice of these parameters are arbitrary. A different choice of these parameters do not bear any effect on the computational complexity of the stronger quantum speed limit bound for mixed quantum states. Next, we obtain the quantum speed limit bound for the mixed quantum states in the same procedure as the other examples mentioned before. We plot our results in Fig.\ref{fig:fighe} From this figure, we again see that our theory give good improvement over the previous MTL quantum speed limit bound and we have $\Delta=\tau_{SQSL}-\tau_{MTL}$ as always positive. The apparent difference in various points can be attributed to the fact that we always choose a random eigenbasis for the calculation of our bound.

\subsection{Two qubit CNOT Hamiltonian}

Two qubit CNOT gate is an important case of a Hamiltonian as this is a part of the universal gates that can be used for performing all sorts of quantum computation. Therefore we choose a Hamiltonian that will represent a two qubit CNOT gate. The form of one such Hamiltonian also called the principal Hamiltonian is given by
$H=\pi \sigma_z^-\otimes \sigma_x^-$
where we have used the following notation
\begin{align}
\sigma_z^\pm=\frac{\mathbb{I}\pm\sigma_z}{2},~~
\sigma_x^\pm=\frac{\mathbb{I}\pm\sigma_x}{2}.
\end{align}
We calculate the quantum speed limit bound for evolution under this Hamiltonian for initially mixed quantum states.

We take the Hilbert space of dimension 4 for our numerical example as represented in subfigures \ref{fig:sfig5a} and \ref{fig:sfig5b} of Fig.\ref{fig:fig5}, i.e., for the case of two qubits. The initial state is taken as the following 
 \begin{align}
 \rho_0=0.7|0\rangle\langle 0|+0.1|1\rangle\langle 1|+0.1|2\rangle\langle 2|+0.1|3\rangle\langle 3|
 \end{align}
 As with all the examples before, we calculate the stronger quantum speed limit bound using the same methods. We check our bound for the above choices of initial mixed quantum state and the Hamiltonian and compare the performance of our optimized bound with the previous bound. The optimization is over $10$ such operators $\sigma$ as in all the above cases. From the figure, we clearly see that we always have $\Delta=\tau_{SQSL}-\tau_{MTL}$ as positive, showing that the stronger quantum speed limit bound derived in this article outperforms the MT like (MTL) bound for mixed quantum states. Also it is natural to expect that our stronger speed limit bound will outperform the MT like bound for mixed quantum states even better when the optimization will be performed over a larger set of $\sigma$.

\subsection{Comparison with other bounds: Perfect state transfer Hamiltonian.}\label{comparison}

Here, we take the example of perfect quantum state transfer for comparing our stronger quantum speed limit bound for mixed quantum states with two other existing important bound for quantum speed limit for mixed quantum states. The Hamiltonian describing this case is given by Eq.(\ref{eqn:pst1}), Eq.(\ref{eqn:pst2}), and the initial quantum state as given by Eq.(\ref{eqn:pst3}). We obtain the quantum speed limit bound for the mixed quantum states in a similar way as before and compare the performance of our optimized bound with the previous two quantum speed limit bound for mixed quantum states as given in \cite{Campaioli2018}. Note that the quantum speed limit bounds given in \cite{Campaioli2018} are better than MT like bounds for most qubit states. We check from the subfigures \ref{fig:sfig6a} and \ref{fig:sfig6b} of Fig.\ref{fig:fig6} that our bound is better than the second and the third existing quantum speed limit bounds as given in \cite{Campaioli2018} in these cases with minimum number of optimizations as stated in their respective figures. The optimization is simple and minimal is completed within about a minute for five optimizations. As a result, this optimization is highly practical and feasible. We notice that the figures \ref{fig:sfig6a} and \ref{fig:sfig6b} look almost identical. As a result, we check whether they are actually numerically identical or their is a difference between them. We plot the difference between the second and third quantum speed limit bounds as given in the paper \cite{Campaioli2018} and plot it in \ref{fig:sfig7a}, which shows that they are actually different by a small margin. Next we check that whether the $+$ and $-$ signs in front of $R(t)$ in our stronger quantum speed limit bounds makes a difference in our stronger quantum speed limit bounds. We again choose the perfect state transfer Hamiltonian as before and plot these bounds as represented in \ref{fig:sfig7b}. As explained in the Fig.\ref{fig:sfig7b}. We see that there are differences with the stronger speed limit bound for plus sign in $R(t)$ with the stronger speed limit bound for minus sign in $R(t)$ for mixed quantum states for the perfect state transfer Hamiltonian from the second and the third previous quantum speed limit bounds as given in the paper \cite{Campaioli2018}. $dp$ represents the difference of our bound with the second (blue) and the third (red) when one uses a $+$ sign in front of $R(t)$ in Eq.(\ref{eqn:meq}) and $dm$ represents the difference of our bound with the second (orange) and the third (green) when one uses a $-$ sign in front of $R(t)$ in Eq.(\ref{eqn:meq}), which highlights all the essential differences between these bounds. This plot also demonstrates that our bound represented by Eq.(\ref{eqn:meq}) performs better than the previous bounds for both the cases of $+$ and $-$ signs in front of $R(t)$.

\section{Conclusions}

In this work, we have derived a stronger quantum speed limit for mixed quantum states using the mixed state generalization of stronger preparation uncertainty relations. We have shown that this bound reduces to that of the pure states under appropriate conditions. Thereafter, we have discussed methods to derive the suitable operators that allows us to calculate our bound. Hereafter we have shown numerically using random Hamiltonians obtained from Gaussian Unitary ensemble that our bound performs better than the mixed state version of the MT bound. The reason for taking random Hamiltonians is nothing but that the technqiue provide valid Hamiltonians that are unlike each other. Also, we have then shown using many suitable analytical examples of Hamiltonians useful in quantum information and computation tasks that the stronger quantum speed limit bound derived here for mixed quantum states also perform better than the MT like bound and also two more existing quantum speed limit bounds for mixed quantum states existing in the current literature. Future directions remain open for comparing our bound to those of other bounds in the literature for mixed quantum states.

\section*{ACKNOWLEDGEMENTS}

S.B. acknowledges discussions with Abhay Srivastav of Harish-Chandra Research Institute, Allahabad, India on an earlier version of the draft of this paper. S. B. acknowledges support from the National Research Foundation of Korea (2020M3E4A1079939, 2022M3K4A1094774) and the KIST institutional program (2E31531). D.T. acknowledges the support from the INFOSYS scholarship and hospitality at Harish-Chandra Research Institute, Allahabad and affiliation of Homi Bhaba National institute during her stay at Harish-Chandra Research Institute. A. K. P. acknowledges the support from the QUEST Grant Q-117 and J C Bose grant from the Department of Science and Technology, India.

\bibliography{ref}

\end{document}